\documentclass[aps,pra,floatfix,nofootinbib]{revtex4}
\usepackage{graphicx}
\usepackage[english]{babel}
\usepackage{amsmath}
\usepackage{amssymb}
\usepackage{ulem}
\usepackage[usenames,dvipsnames]{color}
\usepackage{color}
\definecolor{Blue}{rgb}{0.3,0.3,0.9}
\definecolor{orange}{rgb}{1,0.5,0}
\usepackage{subfigure}
\usepackage{bm}
\usepackage{titlesec}
\newcommand{\out}{\scriptsize\mbox{out}}
\newcommand{\outf}{\scriptsize\mbox{out-fluc}}
\newcommand{\inn}{\scriptsize\mbox{in}}
\newcommand{\innf}{\scriptsize\mbox{in-fluc}}
\newcommand{\sgn}{\mbox{sgn}}
\newcommand{\bx}{{\mathbf x}}
\newcommand{\by}{{\mathbf y}}
\newcommand{\bz}{{\mathbf z}}

\newcommand{\bv}{{\mathbf v}}
\newcommand{\bE}{{\mathbf E}}
\newcommand{\bB}{{\mathbf B}}
\newcommand{\bff}{{\mathbf f}}
\newcommand{\bK}{{\mathbf K}}
\newcommand{\bS}{{\mathbf S}}
\newcommand{\bF}{{\mathbf F}}
\newcommand{\im}{\,{ \rm Im}\, }
\newcommand{\re}{\,{ \rm Re}\, }
\newcommand{\M}{{\scriptsize \mbox{EM}}}

\newcommand{\reg}{\scriptsize\mbox{reg}}
\newcommand{\bG}{{\mathbf G}}

\begin{document}

\title{Spontaneous emission by rotating objects: A {scattering }approach}
\author{Mohammad F. Maghrebi}
\affiliation{Center for Theoretical Physics, Massachusetts Institute of Technology, Cambridge, MA 02139, USA}
\affiliation{Department of Physics, Massachusetts Institute of Technology, Cambridge, MA 02139, USA}
\author{Robert L. Jaffe}
\affiliation{Center for Theoretical Physics, Massachusetts Institute of Technology, Cambridge, MA 02139, USA}
\affiliation{Department of Physics, Massachusetts Institute of Technology, Cambridge, MA 02139, USA}
\author{Mehran Kardar}
\affiliation{Department of Physics, Massachusetts Institute of Technology, Cambridge, MA 02139, USA}

\maketitle

{\bf Abstact}---
  We study the quantum electrodynamics (QED) vacuum in the presence of a
  {body rotating along its axis of symmetry} and show that the object
   spontaneously emits energy if it is lossy.
{The radiated power is expressed as} a general trace formula solely in terms of the scattering matrix, {making} an explicit connection {to} the conjecture of Zel'dovich [JETP Lett. {\bf 14}, 180 (1971)] on rotating objects. We further show that a rotating body drags along nearby objects while making them spin parallel to its own rotation axis.
\\
\\

{Quantum zero-point fluctuations of the electromagnetic field in vacuum lead to macroscopic
manifestations such as the Casimir attraction between neutral conductors~\cite{Casimir48-2}.
When objects are set in motion, they may pull out real photons from
the fluctuating QED vacuum.}
In fact, accelerating boundaries radiate energy, and thus experience friction, through
{the} dynamical Casimir effect~\cite{Moore70}
(see Ref.~\cite{Dalvit11} for a recent review).
Even two parallel plates moving laterally at a constant speed experience a non-contact frictional force~\cite{Pendry97,Volokitin99}.
  While a constant translational motion requires at least two bodies (otherwise, trivial due to Lorentz symmetry), a single spinning object can experience friction. In a recent work~\cite{Manjavacas10}, Manjavacas and Garc\'{i}a  studied {such rotational} friction by {expressing} the polarization fluctuations of a small spinning particle via {the} fluctuation-dissipation theorem, and obtained a frictional force even at zero temperature. In fact, this {problem}
  is closely related to a classical phenomenon known as {\em superradiance} due to Zel'dovich~\cite{Zel'dovich71}. He argues that a rotating object amplifies certain incident waves, and further conjectures that, when quantum {mechanics is considered,} the object should spontaneously emit radiation only for these so-called \emph{superradiating} modes. Indeed this is shown to be the case for a rotating (Kerr) black hole by Unruh~\cite{Unruh74}. This {phenomenon}, however, is different in nature from Hawking {radiation} \cite{Hawking1975}. {One can also find similar effects for a superfluid \cite{Volovik99}.}

In this letter, we treat the vacuum fluctuations in the presence of a rotating object {exactly, except for} the assumption of small velocities to avoid complications of relativity. By incorporating scattering techniques into the Rytov formalism~\cite{Rytov89}, we find a general {trace} formula for the spontaneous emission {by} an arbitrary spinning object, solely in terms of its scattering matrix. We reproduce the results in the literature and find an expression for the radiation {by} a rotating cylinder. Finally, we study the interaction of a rotating body with a \emph{test} object nearby and show that the  rotating body drags along nearby objects while making them rotate parallel to its own rotation axis.

Our starting point is the Rytov formalism~\cite{Rytov89} which relates {fluctuations of the electromagnetic (EM) field to fluctuating \emph{sources} within the material bodies, and in turn to the material's dispersive properties, via the fluctuation-dissipation theorem.}
{The EM} fields are governed by the Maxwell equation
  \begin{equation}
    \left(\nabla \times \nabla \times-\frac{\omega^2}{c^2}\,\epsilon(\omega, \bx)\,\mathbb I\,\right)\,\bE=\frac{\omega^2}{c^2} \bK,
    \label{eq1}
  \end{equation}
  where a linear, and nonmagnetic medium is assumed. Rytov postulates that the sources  undergo fluctuations{,} which are {related to} the imaginary part of the local dielectric response $\epsilon(\omega,\bx)$ {by}
  \begin{equation}\label{Eq: Source fluc}
    \langle  \bK(\omega,\bx) \otimes \bK^*(\omega,\by) \rangle = a_T(\omega)\im \epsilon (\omega,\bx) \delta(\bx-\by)\,\mathbb I.
  \end{equation}
Here, $a_T(\omega))=2\hbar (n_T(\omega)+1/2)$ where $n_T(\omega)=[\exp(\hbar\omega/k_B T)-1]^{-1}$ is the Bose-Einstein occupation number, and $T$ is the temperature.
  {Equations~(\ref{eq1}) and (\ref{Eq: Source fluc})} define the fluctuations of the EM-fields in the presence of a static object at temperature $T$.
  For bodies in uniform motion, {they} are applied in the rest frame of the object, and then transformed to describe the {EM-field} fluctuations in the appropriate laboratory frame.  With all {contributions} of the field correlation functions in a single frame, one can then compute various physical quantities of interest, such as forces, or energy transfer from one object to another, or to the vacuum.
  For non-uniform motion, we assume that the same equations apply locally to the instantaneous rest frame of the body~\cite{Bladel76}.
{This assumption should be valid as long as the rate of acceleration is less than typical
internal frequencies characterizing the object, which are normally quite large.}
The modified Maxwell equations are easier to derive from a Lagrangian,
 $   \mathcal L_\M= \frac{1}{2}\,\epsilon'\,\bE'^2-\frac{1}{2}\,\bB'^2,$
  with the primed fields defined with respect to the {co-moving} reference frame. To incorporate the {fluctuating sources}, we must add
  \begin{equation}
    \Delta \mathcal L= \bK'\cdot\bE'\,
  \end{equation}
  to the Lagrangian, where $\bK'$ is defined in the object's frame. To make contact with the EM-field in the vacuum, we should recast all  fields in the {stationary (lab)} frame.

  In this letter {we assume} that the velocity of the object is small {such that}, to the lowest order in $v/c$, the electromagnetic fields transform as
  \begin{equation} \nonumber
    \bE'=\bE+\frac{\bv}{c}\times \bB, \quad \bB'=\bB -\frac{\bv}{c}\times \bE.
  \end{equation}
  {We discuss below the limits that this assumption places on the generality of our analysis.}
  {In the lab frame  the EM} equation is then given by
  \begin{equation} \label{Eq: EM in moving frame}
    \left[\nabla \times \nabla \times -\frac{\omega^2}{c^2} \mathbb I -\frac{\omega^2}{c^2} \tilde{\mathbb D} (\epsilon'-1) \mathbb D  \right] \bE=\frac{\omega^2}{c^2} \tilde{\mathbb D} \bK',
  \end{equation}
  where
 $   \mathbb D= \mathbb I+\frac{1}{i\omega}{\mathbf v}\times \nabla \times \mbox{and }
    \tilde{\mathbb D}=\mathbb I+\frac{1}{i\omega}\nabla \times {\mathbf v}\times$. Also $\epsilon'$ is the {response function defined in the object's rest frame, now written in terms of laboratory coordinates.}
  Equation~(\ref{Eq: EM in moving frame}) describes the EM-field in the lab frame in terms of  sources {defined} in the object's frame,
$    \langle  \bK'(\omega',\bx') \otimes \bK'^*(\omega',\by') \rangle = a_T(\omega')\im \epsilon (\omega',\bx') \delta(\bx'-\by')\,\mathbb I\,,
$
  where $\omega'$ and $\bx'$ are understood as the frequency and position in the moving frame, which should be transformed to those in the lab frame.

  {Next we turn to} the main point of this study, namely
  {a solid of revolution spinning with angular frequency $\Omega$
  along its axis of symmetry}. We choose time and polar coordinates $(t, r,\phi,z)$ and $(t', r', \phi', z')$ in the lab and the object reference frames respectively, and take the rotation along the $z$ axis. The two coordinate systems are related by
  \begin{equation}\label{Eq: Coordinate transf}
    t'=t, \quad r'=r, \quad \phi'=\phi-\Omega t, \quad z'=z.
  \end{equation}
  Consider a fluctuation of the source {characterized} by frequency $\omega'$ and azimuthal index $m'${,} $\bK'_{\omega', m'}(t',\bx')=e^{-i\omega' t' +i m'\phi'} \bff_{\omega', m'}(r',z')${. Note }that $\bff$ {depends} only on the coordinates $r'$ and $z'$. As a matter of notation, we \emph{define} $\bK(t, \bx)\equiv \bK'(t',\bx')$, \emph{i.e.} we drop the prime when the function is expressed in the lab-frame coordinates. The coordinate transformation in Eq.~(\ref{Eq: Coordinate transf}) then implies that the frequency as seen by the rotating object is shifted by $\Omega m$,
  \begin{equation}\nonumber
    \bK_{\omega,m}(t,\bx)=\bK'_{\omega-\Omega m, m}(t',\bx').
  \end{equation}
  This equation, in turn, recasts the source fluctuations into the lab coordinates:
  \begin{align} \label{Eq: Source fluc moving}
    \langle  & \bK_m(\omega,\bx) \otimes \bK_m^*(\omega,\by) \rangle = a_T(\omega-\Omega m) \times \nonumber \\
    &\times\im \epsilon (\omega-\Omega m, r,z) \frac{\delta(r_\bx-r_\by)\delta(z_\bx-z_\by)}{2\pi r}\,\mathbb I.
  \end{align}
  Here, we used the {constraint} that the object is rotationally symmetric in the $\phi$ direction.
  Equation~(\ref{Eq: EM in moving frame}) together with the {above} equation describes the electromagnetic field fluctuations in the presence of a rotating body.
 {Note that Eqs. (\ref{Eq: Source fluc}) and (\ref{Eq: Source fluc moving}) include both positive and negative frequencies; since $a_T(\omega)$ and $\im \epsilon(\omega)$ are odd functions, rest frame fluctuations are identical for opposite signs.}

 {The electromagnetic fields outside the object receive contributions both from the fluctuating sources within the object and from fluctuations (both zero-point and at finite temperature, thermal) in the vacuum outside the object.}
 The source fluctuations in the vacuum are given by Eq.~(\ref{Eq: Source fluc}) {as described below.}

  {First we consider}  the fluctuations inside the object and find the corresponding field correlation functions. Equation~(\ref{Eq: EM in moving frame}) together with free Maxwell equation in the vacuum ($\epsilon=1$, $\bK'=0$) gives the electric field via the Green's function
  \begin{equation}\label{Eq: E=G K}
    \bE_{\innf} (\omega,\bx) =\frac{\omega^2}{c^2} \int_{\inn} d\bz \, \mathbb G(\omega,\bx,\bz) \cdot\tilde{\mathbb D}{\bK} (\omega,\bz),
  \end{equation}
  where we {have rewritten $\bK'$ in terms of lab-frame coordinates.} The subscript on $\bE$ indicates that the electric field is due to the inside fluctuations (but {is possibly }computed outside the object). The {required} Green's function (with one point inside and the other outside the object) can be formally expanded as
  \begin{align} \label{Eq: G out-in}
    \mathbb G{(\omega,\bx,\bz)}= \frac{i}{2} \sum_{\alpha_m} \bE^{\out}_{\bar\alpha_m}{(\omega,\bx)}\otimes \bF_{\alpha_m}{(\omega,\bz)}.
  \end{align}
  The index $\alpha_m$ {denotes} a set of quantum numbers {including} $m${,} the eigenvalue of the angular momentum along the $z$ direction (in units of $\hbar$). Also $\bar\alpha$ indicates the time reversal of the partial wave $\alpha$. For example, for spherical waves $\alpha_m=(P,l,m)$, where $P$ is the polarization and $l$ is the eigenvalue of the total angular momentum. Here, $\bE^{\out(\inn)}$ is the usual outgoing (incoming) {wave,} while $\bF$ is a solution to the homogeneous EM equation inside the object, \emph{i.e.} Eq.~(\ref{Eq: EM in moving frame}) with the RHS set to zero. The {important constraints are} the continuity equations these functions satisfy on the surface of the object
    \begin{align}\nonumber
    (\bF_{\alpha_{m}})_\|&= \left(\bE^{\inn}_{\alpha_m}+S_{\bar\alpha_{m}} \bE^{\out}_{\alpha_m}\right)_{\|},
  \end{align}
  where $S$ is the scattering matrix. A similar equation holds for the curl acting on the fields. It turns out that the knowledge of the surface value{s  of $\bF$, which in turn can be expressed in terms of the scattering matrix $S_{\alpha_{m}}$, {is}} sufficient for computing the correlation functions.
   Equations~(\ref{Eq: E=G K}) and (\ref{Eq: G out-in}) along with the source fluctuations of Eq.~(\ref{Eq: Source fluc moving}), and a few integrations by parts yield
  \begin{align}\label{Eq: EE Inside fluc}
    \langle &\bE (\omega,\bx) \otimes \bE^* (\omega,\by)\rangle_{\innf}=  \frac{\omega^2 }{4c^2} \sum_{\alpha_m}a_{T}(\omega-\Omega m) \nonumber \\ & \times\left(1-|S_{\alpha_m}|^2\right) \bE^{\out}_{\alpha_m}(\omega,\bx)\otimes \bE^{\out*}_{\alpha_m}(\omega,\by).
  \end{align}

  Next we turn to {fluctuations caused by outside (vacuum) sources}. Equation~(\ref{Eq: Source fluc}) {may appear} to suggest that these {fluctuations} vanish because {$\im \epsilon=0$
in the vacuum, whereas vacuum fluctuations exist even in the absence of any objects.}
{The key is} that an integral over infinite space leads to $1/\im \epsilon$ and thus the limit
{$\im \epsilon\to 0$} should be taken with care \cite{Kruger11}.
{A careful analysis similar to that for the}  inside contribution yields the correlation function due to the outside {sources as}
  \begin{align}\label{Eq: EE Outside fluc}
    \langle \bE (\omega,\bx) \,\otimes\, &\bE^* (\omega,\by)\rangle_{\outf}=  \frac{\omega^2}{4c^2}a_{T_0}(\omega)\times
    \nonumber\\
    \times &\sum_{\alpha_m }\left(\bE^{\inn}_{\alpha_m}(\omega,\bx)+S_{\alpha_m} \bE^{\out}_{\alpha_m}(\omega,\bx)\right)\nonumber \\
    & \otimes \left(\bE^{\inn*}_{\alpha_m}(\omega,\by)+S_{\alpha_m}^* \bE^{\out*}_{\alpha_m}(\omega,\by)\right).
  \end{align}
  Note that the function $a_{T_0}$ is defined at the environment temperature $T_0$ and depends only on $\omega$, unlike the inside fluctuations {which} depend on the shifted frequency, $\omega-\Omega m$.
  {Equation~(\ref{Eq: EE Outside fluc}) has} a rather intuitive form: each term in the dyadic expansion is a linear combination of the incoming wave plus the scattered outgoing wave.

  {Summing Eqs.~(\ref{Eq: EE Inside fluc}) and (\ref{Eq: EE Outside fluc}),} the total correlation function is {given by}
  \begin{equation}\nonumber
    \langle \bE\otimes\bE^*\rangle=\langle \bE\otimes\bE^*\rangle_{\innf}+\langle \bE\otimes\bE^*\rangle_{\outf}.
  \end{equation}
{The above equation completely characterizes the} vacuum fluctuations in the presence of a rotating body possibly {at a different temperature from the vacuum}.
  Interestingly, even at zero temperature, the rotating body spontaneously emits energy, {as}
can be computed by averaging over {the} Poynting vector, $\langle\bE\times \bB\rangle$, and integrating over a surface enclosing the object. The sum of the in- and out-fluctuation contributions yields
  \begin{align}
  \label{eq2}
    P= &\int_{0}^{\infty} \frac{d\omega}{2\pi}  \, \hbar\omega \times \nonumber\\
    &\sum_{\alpha_m} \,[n_T(\omega-\Omega m)-n_{T_0}(\omega)] \,\, \left(1- |S_{\alpha_m}(\omega)|^2\right).
  \end{align}
  Note that the singularity of $n_T$ at $\omega=\Omega m$ is removed since $1-|S|^2$ {vanishes there}.{ Equation~(\ref{eq2})} can be written in a basis-independent form, {where  t}he radiation takes the form
   \begin{align}\label{eq3}
   \!\! P&=
     \int_{0}^{\infty} \frac{d\omega}{2\pi}  \, \hbar\omega \,\times \nonumber\\
    & \mbox{Tr} \left[
    \left(n_T(\omega-\Omega \, \hat l_z)-n_{T_0}(\omega)\right)
    \left(\mathbb I- \mathbb S^\dagger(\omega)\mathbb S (\omega)\right)\right].
  \end{align}
{Here} $\hat l_z$ is the $z$-component of the angular momentum operator (in units of $\hbar$) and $\mathbb S$ is the (basis-independent) scattering matrix. Note that this equation reduces to the heat radiation from a static object in the limit of zero angular velocity~\cite{Beenakker98,Kruger11}. We are specifically interested in zero temperature. In this limit, the number of radiated photons in the partial wave $\alpha_m$ is
  \begin{align}\label{Eq: E Radiation}
    \frac{d{\cal N}_{\alpha_m}}{d\omega}&= \Theta(\Omega m-\omega)\, (|S_{\alpha_m}(\omega)|^2-1),
  \end{align}
  where $\Theta$ is the Heaviside function.
  Hence, the radiation only comes from the frequency window $0 <\omega  <\Omega m$ for the $m^{\rm th}$ partial wave. In fact, Zel'dovich {proposed} \emph{classical} superradiance for the same frequency regime~\cite{Zel'dovich71}. He considered a rotating cylinder and argued that exactly for waves in the above {frequency range}, the amplitude of the scattered wave (in absolute value) is larger than one, \emph{i.e.} $|S_{\alpha_m}(\omega)|>1$; see Refs.~\cite{Bekenstein98,Richartz09} for further discussion. He further conjectured that taking quantum mechanics into account {would }lead to spontaneous emission. Equations~(\ref{eq2}-\ref{Eq: E Radiation}) are, to our knowledge, the first expressions for spontaneous emission by a rotating object given explicitly in terms of the scattering matrix, which show that, at zero temperature, the radiation is generated exactly in the superradiating channels.

{Note that, having assumed non-relativistic velocities,} we must keep only the leading contribution in $\Omega R/c$ to radiation from each partial wave. In fact, higher partial waves typically contribute in higher powers of this quantity. Therefore, to compute the total radiation at zero temperature, {we} must keep only the lowest partial wave, while at finite temperature higher partial waves can make a comparable, or even larger, contribution due to {the Boltzmann weight}.

{Using the} general expression for radiation, we {now} discuss some {simple}
special cases, {namely} a sphere and a cylinder.  {To find the $S$-matrix of a rotating object we have to solve a complicated equation --- Eq.~(\ref{Eq: EM in moving frame}) with the RHS set to zero, but the task is made easier by making a further assumption that the object's radius is small enough so that $|\sqrt{\epsilon}| \Omega R/c \ll 1$, which allows us to neglect the explicit dependence on the velocity ($\mathbb D\approx \mathbb I$).  Also, for small $\Omega$ we need only consider the low-frequency response.}

For a \emph{sphere} the $T$-matrix is {related to its} (electric) polarizability $\alpha$ { by:}
 $T^{EE}_{1m 1m}(\omega)=i \frac{2\omega^3}{3c^3}\alpha(\omega-\Omega m)$; the argument of $\alpha$ is $\omega-\Omega m$ because that of $\epsilon$ is shifted.
 {The value of these functions for negative arguments can be obtained through their analytic properties.} The scattering matrix is {related to the $T$-matrix via }$S=1+2T$, and the energy radiation in the lowest partial wave can be computed by Eq.~(\ref{eq2}). The result is indeed in agreement with Ref.~\cite{Manjavacas10}.

  The $T$-matrix for the \emph{cylinder} is rather complicated. Further, one must take into account all polarizations.
  Here, we quote the final result for a slowly rotating cylinder of radius $R$ and length $L$, and at zero temperature everywhere,
  \begin{align}
    P&  =\frac {2\hbar L R^2}{3\pi c^3}\int_{0}^{\Omega} d\omega \, \omega^4 \left|\im \frac{\epsilon(\omega-\Omega)-1}{\epsilon(\omega-\Omega)+1}\right|.
  \end{align}
{This equation, valid for arbitrary $\epsilon$, reproduces the result in Ref.~\cite{Zel'dovich86} in the limit of small conductivity. }

  The radiation from a rotating object  {exerts} pressure on nearby objects. Consider a spherical object {with angular velocity $\Omega$} {along the $z$ direction} {and} a second small spherical {body}{--- a \emph{test {object}} --- } at rest, placed at a separation $d$ {on} the $x$ axis. One finds that, in addition to the force along the $x$ axis, there is a tangential force in the $y$ direction. Furthermore, a torque {is} exerted on the test object. To compute this effect, we break up the EM-field correlation function into radiation (due to propagating photons) and non-radiation (due to zero-point fluctuations) parts:
  \begin{equation}\nonumber
    \langle \bE \otimes \bE^* \rangle =    \langle \bE \otimes \bE^* \rangle_{\rm rad} +    \langle \bE \otimes \bE^* \rangle_{\rm non-rad}\,.
  \end{equation}
  {The non-radiation part gives the Casimir force which makes no contribution to the tangential force or the torque.} At zero temperature, the radiation comes only through the superradiating modes {with} $\omega<\Omega m$,
  \begin{align}\label{?}
    \langle &\bE (\omega,\bx) \otimes \bE^* (\omega,\by)\rangle_{\rm rad}=  \frac{\hbar  \omega^2}{2c^2} \sum_{\alpha_m}\Theta(\Omega m-\omega) \nonumber \\ & \times\left(|S_{\alpha_m}|^2-1\right) \bE^{\out}_{\alpha_m}(\omega,\bx)\otimes \bE^{\out*}_{\alpha_m}(\omega,\by).
  \end{align}
  We assume that $d$ is large compared to the length scales of both the rotating body and the test object, and thus compute the first reflection of the radiation off of the test object~\cite{Maghrebi10-2}. First we transform the outgoing waves to regular waves about the origin of the second object via translation matrices~\cite{Rahi09}. {These waves} scatter on the second object giving
  \begin{equation}
    M= \frac{\hbar c^2}{8\pi d^2} \int_{0}^{\Omega} d\omega \, \frac{1}{\omega^2} (|S_{11E}|^2-1)(1-|\mathfrak{S}_{11E}|^2),
  \end{equation}
  for the torque, and
  \begin{equation}
    F_y=\frac{\hbar}{32\pi d}\int_{0}^{\Omega} d\omega \,(|S_{11E}|^2-1) (1-\re\mathfrak{S}_{11E}),
  \end{equation}
  for the shear force on the test object, where $S_{11E}$ and $\mathfrak{S}_{11E}$ are the scattering matrices of the rotating and the test object, respectively, in the lowest partial wave ($l=m=1, P=E$).
  This partial wave gives the leading order in $1/d$ for small $\Omega$.
  Note that the torque falls off as $1/d^2$ with separation while the force goes as $1/d$. Furthermore the signs are positive, \emph{i.e.} a rotating body drags along objects nearby while making them rotate parallel to its own rotation axis.

  To get an estimate for {the magnitude of} radiation effects, {we} consider a {rapidly} spinning nanotube of radius $R$ and length $L$, and assume that $\Omega R/c$ is small. We then find that the rotation slows down by an order of magnitude over a time scale of $\tau\sim (I/\hbar ) \, {(}c^3 /L R^2\Omega^3{)}$.
  The moment of inertia of a nanotube can be as small as $10^{-33}$ in SI units~\cite{Sadeghpour02} (compare with $\hbar \approx 10^{-34}$). So even at small velocities, $\tau$ can be of the order of a few hours.

  {We have derived a} universal formula for the spontaneous emission by a rotating object in terms of the scattering matrix {that} makes an explicit connection {to} the physical principles behind superradiance. Furthermore, it allows one to circumvent assumptions about small size or small conductivity used in the literature. We also believe the current formalism naturally generalizes to relativistic motion. Finally, generalization of the technical and conceptual aspects of this work to the interaction of multiple (moving) objects would be worthwhile.

  We thank G. Bimonte, T.~Emig, V.~Golyk, N.~Graham, M.~Kr{\"{u}}ger, and H.~Reid for helpful conversations.
  MFM is indebted to M.~Kr{\"{u}}ger for many discussions, and to T. Emig for providing the $T$-matrix of a cylinder.
  This work was supported by the U.\ S.\ Department of Energy under cooperative research agreement \#DF-FC02-94ER40818 (MFM and RLJ), {and} NSF Grant No. DMR-08-03315 (MK).


\newpage
{\begin{center}
\large \bf  Supplementary Material
\end{center} }
\setcounter{equation}{0}
 \renewcommand{\theequation}{S\arabic{equation}}

\section{Static object in thermal non-equilibrium}

We start with a static object at a finite temperature different from that of the environment. The generalization to moving objects closely follows the techniques developed in this section.

The Rytov formalism \cite{Rytov89,Kruger11} for electromagnetism (EM) is formulated in terms of fluctuating sources $\bK$
\begin{align}
  \begin{cases}
    \nabla \times \bE=i \frac{\omega}{c} \bB, & \\
    \nabla \times \bB=-i\,\epsilon(\omega) \, \frac{\omega}{c} \bE-i\frac{\omega }{c}\bK,  &
  \end{cases}
\end{align}
or equivalently
\begin{equation}\label{Eq: EM eqn}
 \left( \nabla \times \nabla \times -\, \epsilon(\omega)\, \frac{\omega^2}{c^2} \mathbb I \right)\bE= \frac{\omega^2}{c^2} \bK.
\end{equation}
The source fluctuations are related to the imaginary part of the local dielectric function
\begin{equation}\label{Eq: Source Fluc}
  \langle \bK(\omega,\bx) \otimes \bK^*(\omega,\by)\rangle =a(\omega) \im \epsilon (\omega, \bx) \delta(\bx-\by) \mathbb,
\end{equation}
where
\begin{equation}
  a_T(\omega)=2\hbar \left(n_T(\omega)+\frac{1}{2}\right), \hskip .3in \mbox{with} \hskip .1in  n_T(\omega)= \frac{1}{\exp(\hbar\omega/k_B T)-1}.
\end{equation}
The electric field is obtained by the current via the Green's function, $\bE=\frac{\omega^2}{c^2}\int \mathbb G \bK$. We are interested in the EM field fluctuations outside the object in order to compute, for example, the Poynting vector. The field fluctuations receive contributions both from the fluctuating sources within the object and from fluctuations (zero-point and at finite temperature, thermal) in the vacuum outside the object.
We first consider the EM field fluctuations due to source fluctuations outside the object.

The (dyadic) electromagnetic Green's function is defined by
\begin{equation}
  (\nabla \times \nabla \times - \epsilon(\omega)\, \frac{\omega^2}{c^2}\mathbb I \, )\,\mathbb G(\omega,\bx,\bz)= \mathbb I \, \delta (\bx-\bz).
\end{equation}
In an appropriate coordinate system $(\xi_1, \xi_2, \xi_3)$ the free Green's function can be broken up along the coordinate $\xi_1$. In the absence of the object, {\it i.e.} in empty space, the (free) Green's function can be written as \cite{Rahi09}
\begin{align}\label{Eq: G out-out0}
\mathbb G(\omega; \bx,\bz)=i
\begin{cases}
 \sum_\alpha  {\mathbf E}^{\out}_{\bar \alpha}(\omega,\bx)  \otimes {\mathbf E}^{\reg}_{\alpha}(\omega,\bz) & \xi_1(\bx)> \xi_1(\bz), \\
\sum_\alpha {\mathbf E}^{\reg}_{\alpha}(\omega,\bx) \otimes {\mathbf E}^{\out}_{\bar \alpha}(\omega,\bz) & \xi_1(\bx)< \xi_1(\bz),
\end{cases}
\end{align}
where $\bE^{\out}$ is the outgoing electric filed normalized such that the corresponding (outgoing) current (energy flux) is $\omega$. Also $\bE^{\reg}$ defines a solution to the EM field regular everywhere in the space. The index $\alpha$ runs over partial waves; $\bar\alpha$ indicates the partial wave which is related to $\alpha$ by time reversal symmetry.
In the presence of an external object, the free Green's function should be modified to incorporate the scattering from the object (with $\xi_1(\bx)< \xi_1(\bz)$)
\begin{align}\label{Eq: G out-out}
  \mathbb G(\omega,\bx,\bz)= \frac{i}{2}\sum_\alpha \left(\bE^{\inn}_\alpha(\omega,\bx)+S_\alpha \bE^{\out}_\alpha(\omega,\bx)\right)\otimes \bE^{\out}_{\bar\alpha}(\omega,\bz) ,
\end{align}
where $S_\alpha$ is the scattering matrix\footnote{We have assumed that the scattering matrix is diagonal in partial waves by choosing an appropriate coordinate system, for example by taking $\xi_1$ to be constant on the object's surface.} corresponding to the partial wave $\alpha$ and $\bE^{\inn}$ is the incoming wave whose current is normalized to negative $\omega$. In the absence of the object, $S_\alpha=1$ and Eq.~(\ref{Eq: G out-out}) reduces to the free Green's function ($(\bE^{\inn}_\alpha+ \bE^{\out}_\alpha)/2=\bE^{\reg}_\alpha$). {\it Note that the Green's function in the last equation is defined with both points outside the object.}

The EM correlation function due the outside source fluctuations is then given by
\begin{align}
  \langle \bE (\omega,\bx) \otimes \bE^* (\omega,\by)\rangle_{\outf} &= \frac{\omega^4}{c^4} \int d\bz\langle \mathbb G(\omega, \bx,\bz) \bK (\omega,\bz)\otimes\mathbb G^*(\omega,\by,\bz) \bK^* (\omega,\bz)\rangle \\
  &= a_{T_0}(\omega) \frac{\omega^4}{c^4} \, \im \epsilon_D \int_{\out}  d\bz \, \mathbb G(\omega,\bx,\bz) \cdot \mathbb G^*(\omega,\by,\bz),
\end{align}
where the dot product should be understood as the contraction of the second subindex of the two dyadic functions. The last line in this equation is obtained according to Eq.~(\ref{Eq: Source Fluc}). Here $T_0$ is the temperature of the environment and $\epsilon_D$ is the dielectric function of the vacuum. The latter can be set to one only in the end for the reason explained in the letter: the integral over infinite space may bring down a factor of 1/$\im \epsilon_D$ \cite{Kruger11}.
However, the integration over any finite region vanishes as we take the limit $\im \epsilon_D \to 0$. Therefore we can choose the domain of integration over $\bz$ such that $\xi_1(\bz)>\xi_1(\bx),\xi_1(\by)$. This allows us to use the partial wave expansion of the Green's function in Eq.~(\ref{Eq: G out-out}). We then find
\begin{align}
\langle \bE (\omega,\bx) \otimes \bE^* (\omega,\by)\rangle_{\outf}=  &\frac{\omega^4}{4c^4}\, a_{T_0}(\omega)\sum_{\alpha, \beta }\left(\bE^{\inn}_\alpha(\omega,\bx)+S_\alpha \bE^{\out}_\alpha(\omega,\bx)\right)\otimes \left(\bE^{\inn*}_\beta(\omega,\by)+S_\beta^* \bE^{\out*}_\beta(\omega,\by)\right) \times \nonumber \\
&\times (\im \epsilon_D)\int_{\out} d\bz \, \bE_{\bar\alpha}^{\out}(\omega,\bz)\cdot \bE_{\bar\beta}^{\out*}(\omega,\bz).
\end{align}
Here and in subsequent parts, we frequently compute volume integrals similar to the last line of this equation, which we write as
\begin{align}
  &(\im \epsilon_D)\int_{\out} d\bz \, \bE_{\bar\alpha}^{\out}(\omega,\bz)\cdot \bE_{\bar\beta}^{\out*}(\omega,\bz)\nonumber\\
  =&\frac{1}{2i}\int_{\out} d\bz \, \left[\epsilon_D \,\bE_{\bar\alpha}^{\out}(\omega,\bz)\cdot \bE_{\bar\beta}^{\out*}(\omega,\bz)-\bE_{\bar\alpha}^{\out}(\omega,\bz)\cdot \epsilon_D^* \,\bE_{\bar\beta}^{\out*}(\omega,\bz)\right]\nonumber\\
  =&\frac{c^2}{2i\omega^2} \int_{\out} d\bz \, \left[(\nabla \times \nabla \times \,\bE_{\bar\alpha}^{\out}(\omega,\bz))\cdot \bE_{\bar\beta}^{\out*}(\omega,\bz)-\bE_{\bar\alpha}^{\out}(\omega,\bz)\cdot \nabla \times \nabla \times \,\bE_{\bar\beta}^{\out*}(\omega,\bz)\right]\nonumber,
\end{align}
where in the last line we have used the homogenous version of Eq.~(\ref{Eq: EM eqn}) (with the RHS set to zero). The volume integration can be then recast as a surface integral. The boundary comprises of two surface, one at the infinity and the other at a finite distance from the object. The infinitesimal imaginary part of the dielectric function guarantees that outgoing functions are exponentially decaying at very large distances and thus the surface integral at the infinity does not contribute. Therefore we obtain
\begin{align}
\im \epsilon_D\int_{\out} d\bz \, \bE_{\bar\alpha}^{\out}(\omega,\bz)\cdot \bE_{\bar\beta}^{\out*}(\omega,\bz)
&=  \frac{i c^2}{2\omega^2} \oint d{\mathbf \Sigma}\cdot \left[(\nabla \times \bE^{\out}_{\bar\alpha }(\omega,\bz))\times \bE^{\out*}_{\bar\beta}(\omega,\bz)+\bE^{\out}_{\bar\alpha }(\omega,\bz)\times \nabla \times \bE^{\out*}_{\bar\beta}(\omega,\bz)\right].
\end{align}
A vector form of Green's theorem can be used to prove, consistent with our normalization, that (see Appendix A)
\begin{align}\label{Eq: identity}
  \frac{i}{2}\oint d{\mathbf \Sigma}\cdot \left[(\nabla \times \bE^{\out/\inn}_\alpha (\omega,\bz))\times \bE^{\out/\inn*}_{\beta}(\omega,\bz)+\bE^{\out / \inn}_\alpha (\omega,\bz)\times \nabla \times \bE^{\out/\inn*}_{\beta}(\omega,\bz)\right]=\pm\delta_{\alpha \beta}.
\end{align}
(Also, $\oint d{\mathbf \Sigma}\cdot \left[(\nabla \times \bE^{\out/\inn}_\alpha (\omega,\bz))\times \bE^{\inn/\out*}_{\beta}(\omega,\bz)+\bE^{\out / \inn}_\alpha (\omega,\bz)\times \nabla \times \bE^{\inn/\out*}_{\beta}(\omega,\bz)\right]=0.
$)

Therefore, the correlation function of the EM fields takes the form
\begin{align}\label{Eq: <EE> out static}
\langle \bE (\omega,\bx) \otimes \bE^* (\omega,\by)\rangle_{\outf}=   a_{T_0}(\omega)\frac{\omega^2 }{4c^2} \sum_{\alpha} \left(\bE^{\inn}_\alpha(\omega,\bx)+S_\alpha \bE^{\out}_\alpha(\omega,\bx)\right)\otimes \left(\bE^{\inn*}_\alpha(\omega,\by)+S_\alpha^* \bE^{\out*}_\alpha(\omega,\by)\right).
\end{align}
The Poynting vector, $\bS=c\, \bE \times \bB$, gives the rate of energy radiation. So the radiation due to the outside fluctuations is given by
\begin{align}
 P_{\outf}&= \int_{0}^{\infty} \frac{d\omega}{2\pi}\oint d{\mathbf \Sigma} \cdot \frac{ic^2}{ \omega} \langle (\nabla \times \bE)\times \bE^* +\bE \times \nabla \times \bE^* \rangle_{\outf} \nonumber \\
 &= \int_{0}^{\infty} \frac{d\omega }{2\pi}\, \sum_{\alpha}  (-1+ |S_\alpha(\omega)|^2) \,a_{T_0}(\omega)\frac{i\omega}{4}\oint d{\mathbf \Sigma} \cdot \left[(\nabla \times \bE^{\out}_\alpha)\times \bE_\alpha^{\out*} +\bE^{\out}_\alpha \times \nabla \times \bE_\alpha^{\out *} \right] \nonumber \\
  &=    \frac{1}{4\pi}\int_{0}^{\infty} d\omega \, \omega \,a_{T_0}(\omega)\, \sum_{\alpha} (-1+ |S_\alpha(\omega)|^2).
\end{align}
In the last line, we used Eq.~(\ref{Eq: identity}).

The field correlation function induced by the inside fluctuations can be computed similarly. In this case, however, the Green's function with one point inside the object should be used.
First note that the two points do not coincide so the Green's function satisfies a {\it homogeneous} equation inside with respect to the {\it smaller} coordinate while it satisfies the free EM equation outside the object with respect to the {\it larger} coordinate. Hence, we can expand the Green's function as (with $\xi_1(\bz)<\xi_1(\bx)$)
\begin{equation} \label{Eq: Green fn in-out 0}
  \mathbb G(\omega, \bx,\bz)=  \frac{i}{2}\sum_{\alpha}  \left( A \, \bE^{\out}_{\bar\alpha}(\omega,\bx) +B \, \bE^{\inn}_{\bar\alpha}(\omega,\bx)\right)\otimes \bF_\alpha(\omega,\bz).
\end{equation}
The prefactor ${i}/{2}$ is chosen for convenience, $A$ and $B$ are constants to be determined, and $\bF_\alpha$ is defined as a solution to the EM equation inside the object
\begin{equation}
\left(\nabla \times \nabla \times- \, \epsilon(\omega, \bx)\, \omega^2 \mathbb I\right) \bF_\alpha(\omega,\bx)=0.
\end{equation}
We can determine the coefficients $A$ and $B$ and the normalization of $\bF$ by matching the Green's functions approaching a point on the boundary from inside and outside the object
\begin{equation}
 \mathbb G(\omega, \bx,\by){\mid}_{\by\to \Sigma^{-}}=  \mathbb G(\omega, \bx,\by){\mid}_{\by \to \Sigma^+},  \hskip .2in \mbox{} \xi_1(\bx)>\xi_1(\by),
\end{equation}
where $\Sigma$ represents the boundary.
Comparing the two Green's functions given by Eqs. (\ref{Eq: G out-out}) and (\ref{Eq: Green fn in-out 0}), we find ($A=1, B=0$)
\begin{align}
  \mathbb G(\omega, \bx,\bz)= \frac{i}{2}\sum_{\alpha}  \bE^{\out}_{\bar\alpha}(\omega,\bx)\otimes \bF_\alpha(\omega, \bz),
\end{align}
where $\bF$ is normalized by the continuity of the Green's function which implies matching boundary conditions for the parallel components of electric and magnetic fields (the latter because $\mu=1$)
\begin{align}\label{Eq: Boundary condition}
  \bF_\alpha(\omega, \bz)_\|&= \left(\bE^{\inn}_\alpha(\omega,\bz)+S_\alpha \bE^{\out}_\alpha(\omega,\bz)\right)_{\|}, \nonumber\\
    \left(\nabla \times \bF_\alpha(\omega,\bz)\right)_\|&= \left(\nabla \times \bE^{\inn}_\alpha(\omega,\bz)+S_\alpha \nabla \times \bE^{\out}_\alpha(\omega,\bz)\right)_{\|}.
\end{align}

The correlation function due to the inside fluctuations is then given by
\begin{align}
\langle \bE (\omega,\bx) \otimes \bE^* (\omega,\by)\rangle_{\innf}=  &a_{T}(\omega)\,\frac{\omega^4}{4c^4} \sum_{\alpha, \beta }\bE^{\out}_{\bar\alpha}(\omega,\bx)\otimes \bE^{\out*}_{\bar\beta}(\omega,\by)  \int_{\inn} d\bz \,\bF_{\alpha}(\bz)\cdot \im \epsilon(\omega, \bz)\,\bF_{\beta}^{*}(\bz).
\end{align}
In this equation, $T$ is the object's temperature. Again using the wave equation for $\bF$, the volume integral can be cast as a surface term
\begin{equation}
  \int_{\inn} d\bz \, \bF_{\alpha}(\bz)\cdot \im \epsilon(\omega, \bz)\,\bF_{\beta}^{*}(\omega,\bz) = \frac{c^2}{2i\omega^2} \oint d{\mathbf\Sigma}\cdot  \left[\left(\nabla \times \bF_\alpha (\omega,\bz)\right)\times \bF^{*}_{\beta}(\omega, \bz)+\bF_\alpha (\omega,\bz)\times \nabla \times \bF^{*}_{\beta}(\omega,\bz)\right].
\end{equation}
The continuity equations can be used to evaluate the surface integral
\begin{equation}\label{Eq: Integral over F: static}
  \int_{\inn} d\bz \, \bF_{\alpha}(\omega,\bz)\cdot \im \epsilon(\omega, \bz)\,\bF_{\beta}^{*}(\omega,\bz) =  \frac{c^2}{\omega^2}\delta_{\alpha\beta} \left(1-|S_\alpha|^2\right).
\end{equation}
The field correlation function then becomes\footnote{We have changed $\alpha\ \to \bar\alpha$; note that $|S_{\bar\alpha}|=|S_{\alpha}|$ due to time reversal symmetry.}
\begin{align}
\langle \bE (\omega,\bx) \otimes \bE^* (\omega,\by)\rangle_{\innf}=  & a_{T}(\omega)\frac{\omega^2}{4 c^2} \sum_{\alpha}\left(1-|S_\alpha|^2\right) \bE^{\out}_{\alpha}(\omega,\bx)\otimes \bE^{\out*}_{\alpha}(\omega,\by).
\end{align}
The radiation power due to the inside fluctuations can be computed similar to that of the outside fluctuations
\begin{align}
 P_{\innf}&= \frac{1}{4\pi}\int_{0}^{\infty} d\omega \, \omega \,a_{T}(\omega)\, \sum_{\alpha} (1- |S_\alpha(\omega)|^2).
\end{align}
The total radiation power then becomes
\begin{align}
  P&= \frac{1}{4\pi}\int_{0}^{\infty} d\omega \, \omega \,\left(a_{T}(\omega)-a_{T_0}(\omega)\right)\, \sum_{\alpha} (1- |S_\alpha(\omega)|^2) \nonumber \\
  & =\int_{0}^{\infty} \frac{d\omega}{2\pi} \, \hbar\omega \,\left(n_{T}(\omega)-n_{T_0}(\omega)\right)\, \sum_{\alpha} (1- |S_\alpha(\omega)|^2).
\end{align}
In the last line, the radiation is given in terms of the occupation number.
\section{Moving object in thermal and dynamical non-equilibrium}
The EM wave equation for a moving medium can be inferred from a Lagrangian. A dielectric object is described by
\begin{align}
  \mathcal L
   = \frac{1}{2} \epsilon'\, \bE'^2 - \frac{1}{2} \bB'^2
\end{align}
where $\bE'$ and $\bB'$ are the EM fields in the comoving frame which relate to the EM fields in the lab frame via
\begin{equation}\label{Eq: rel transf}
  \bE'=\bE+{\frac{\mathbf v}{c}}\times \bB, \hskip .2in   \bB'=\bB-{\frac{\mathbf v}{c}}\times \bE,
\end{equation}
to the lowest order in velocity. Also $\epsilon'=\epsilon (\omega',\bx')$ is the dielectric function given in the coordinates of the moving object.
The wave equation for the electric field becomes\footnote{The Lagrangian can be cast as
\begin{equation*}
    \mathcal L
   = \mathcal L_0 +\frac{1}{2} (\epsilon'-1)\, \bE'^2,
\end{equation*}
where $\mathcal L_0=\frac{1}{2} \, \bE'^2 - \frac{1}{2} \bB'^2$. Notice that $\mathcal L_0$ is invariant under the transformation in Eq.~(\ref{Eq: rel transf}) to the first order in $v/c$, {\it i.e.} $\mathcal L_0=\frac{1}{2} \, \bE^2 - \frac{1}{2} \bB^2 +\mathcal O(v^2/c^2)$. Therefore, the Lagrangian takes the form
\begin{align*}
    \mathcal L
   &= \frac{1}{2} \, \bE^2 - \frac{1}{2} \bB^2+\frac{1}{2} (\epsilon'-1)\, \bE'^2 +\mathcal O(v^2/c^2) \\
   &= \frac{1}{2} \, |\bE|^2 - \frac{c^2}{2 \omega^2} |\nabla \times \bE|^2+\frac{1}{2} (\epsilon'-1)\, |\mathbb D \,\bE|^2+\mathcal O(v^2/c^2).
\end{align*}
Equation (\ref{Eq: EM eqn moving}) follows straightforwardly.
}
\begin{equation}\label{Eq: EM eqn moving}
  \left[\nabla \times \nabla \times -\frac{\omega^2}{c^2} \mathbb I -\frac{\omega^2}{c^2} \tilde{\mathbb D} (\epsilon'-1) \mathbb D  \right] \bE=0,
\end{equation}
where
\begin{align}
  \mathbb D= \mathbb I+\frac{1}{i\omega}{\mathbf v}\times \nabla \times\,, \hskip .2in
  \tilde{\mathbb D}= \mathbb I+\frac{1}{i\omega}\nabla \times{\mathbf v}\times.
\end{align}
The coupling with the (fluctuating) currents can be formulated by adding
\begin{equation}
  \Delta\mathcal L= \bK'\cdot\bE'
\end{equation}
to the Lagrangian.
The inhomogeneous EM equation including the sources takes the form
\begin{equation}
  \left[\nabla \times \nabla \times -\frac{\omega^2}{c^2} \mathbb I -\frac{\omega^2}{c^2} \tilde{\mathbb D} (\epsilon'-1) \mathbb D  \right] \bE=\frac{\omega^2}{c^2} \tilde{\mathbb D} \bK'.
\end{equation}

The motion of the object modifies the source fluctuations inside the object, however they affect the outside (vacuum) fluctuations only through a modified scattering matrix appropriately computed for a moving object. Therefore the EM field correlation function due to the latter fluctuations are given similarly to that of Eq.~(\ref{Eq: <EE> out static}),
\begin{align}\label{Eq: <EE> out moving}
\langle \bE (\omega,\bx) \otimes \bE^* (\omega,\by)\rangle_{\outf}=   a_{T_0}(\omega)\frac{\omega^2 }{4c^2} \sum_{\alpha_m}\left(\bE^{\inn}_{\alpha_m}(\omega,\bx)+S_{\alpha_m} \bE^{\out}_{\alpha_m}(\omega,\bx)\right)\otimes \left(\bE^{\inn*}_{\alpha_m}(\omega,\by)+S_{\alpha_m}^* \bE^{\out*}_{\alpha_m}(\omega,\by)\right).
\end{align}
Note that the $S$-matrix describes scattering from a moving object. The partial-wave index $\alpha_m$ includes $m$, the eigenvalue of the angular momentum along the $z$-direction (in units of $\hbar$); this notation proves useful in the following.

The inside fluctuations, on the other hand, are defined with respect to the rest frame of the object, that is,
\begin{equation}
  \langle \bK'(\omega',\bx') \otimes {\bK'}^*(\omega', \by')\rangle = a_{T}(\omega') \im \epsilon (\omega', \bx') \delta(\bx'-\by')\mathbb I.
\end{equation}
Defining $    \bK_{\omega m}(t,\bx)=\bK'_{\omega' m'}(t',\bx')$ with $t$ and $\bx$ being the time and space coordinates in the lab frame, one can see that
\begin{equation}
  \begin{cases}
    m'=m, & \\
    \omega'=\omega-\Omega m. &
  \end{cases}
\end{equation}
This modifies the spectral density of source fluctuations merely through replacing the argument of $\epsilon$ and $a_T$ by $\omega-\Omega m$. Therefore, the inside source fluctuations from the point of view of the lab-frame observer are given by
  \begin{align} \label{Eq: Source fluc moving}
    \langle  & \bK_m(\omega,\bx) \otimes \bK_m^*(\omega,\by) \rangle = a_T(\omega-\Omega m) \im \epsilon (\omega-\Omega m, r,z) \frac{\delta(r_\bx-r_\by)\delta(z_\bx-z_\by)}{2\pi r}\,\mathbb I.
  \end{align}
Hereafter, we use the same notation $\mathbb G$ for the Green's function although we keep in mind that the latter is the Green's function in the presence of a moving object described by Eq.~(\ref{Eq: EM eqn moving}).
The EM field correlation function is
\begin{align}
  \langle \bE (\omega,\bx) \otimes \bE^* (\omega,\by)\rangle_{\innf} &= \frac{\omega^4}{c^4} \int_{\inn} d\bz\,\langle \mathbb G(\omega,\bx,\bz) \tilde{\mathbb D}{\bK} (\omega,\bz)\cdot\mathbb G^*(\omega,\by,\bz) {\tilde{\mathbb D}}^* {\bK}^* (\omega,\bz)\rangle.
\end{align}
We can expand the Green's function similar to the previous section
\begin{align}
  \mathbb G(\omega,\bx,\bz)= \frac{i}{2}\sum_{\alpha_m}  \bE^{\out}_{\bar \alpha_m}(\omega,\bx)\otimes \bF_{\alpha_m}(\omega,\bz),
\end{align}
where $\bF$ is a solution to the {\it modified} EM equation inside the dielectric object
\begin{equation}\label{rel em inside sln}
  \left[\nabla \times \nabla \times -\omega^2 \mathbb I -\omega^2 \tilde{\mathbb D} (\epsilon'-1) \mathbb D  \right] \bF=0,
\end{equation}
and satisfies boundary conditions similar to Eq.~(\ref{Eq: Boundary condition}) only with the scattering matrices understood for a moving object\footnote{Note that the scattering matrix is given for $\bar\alpha$. This is because the Green's function in the presence of a moving object is no longer symmetric with respect to its spatial arguments but satisfies a rather different symmetry. If Eq.~(\ref{Eq: G out-out}) defines the Green's function with $\xi_1(\bx)< \xi_1(\bz)$, then
\begin{align*}
  \mathbb G(\omega,\bz,\bx)= \frac{i}{2}\sum_\alpha \bE^{\out}_{\bar\alpha}(\omega,\bz)\otimes\left(\bE^{\inn}_\alpha(\omega,\bx)+S_{\bar\alpha} \bE^{\out}_\alpha(\omega,\bx)\right).
\end{align*}
}
\begin{align}\label{Eq: Boundary condition2}
  \bF_{\alpha_m}(\omega, \bz)_\|&= \left(\bE^{\inn}_{\alpha_{m}}(\omega,\bz)+S_{\bar\alpha_{m}} \bE^{\out}_{\alpha_{m}}(\omega,\bz)\right)_{\|}, \nonumber\\
    \left(\nabla \times \bF_{\alpha_m}(\omega,\bz)\right)_\|&= \left(\nabla \times \bE^{\inn}_{\alpha_{m}}(\omega,\bz)+S_{\bar \alpha_{m}} \nabla \times \bE^{\out}_{\alpha_{m}}(\omega,\bz)\right)_{\|}.
\end{align}
The correlation function of the electric fields is then given by
\begin{align}
\langle \bE (\omega,\bx) \otimes \bE^* (\omega,\by)\rangle_{\innf}=  \, &\frac{\omega^4}{4c^4} \sum_{\alpha_{m}, \beta_m } a_{T}(\omega-\Omega m)\,\bE^{\out}_{\bar\alpha_m}(\omega,\bx)\otimes \bE^{\out*}_{\bar\beta_m}(\omega,\by) \times \nonumber \\
&\times \int_{\inn} d\bz \, \mathbb D\,\bF_{\alpha_m}(\omega,\bz)\cdot \im \epsilon(\omega-\Omega m, \bz)\,{\mathbb D}^*\bF_{\beta_m}^{*}(\omega,\bz).
\end{align}
The volume integral can be computed similar to that of the previous section. We write the second line of the last equation as ($\epsilon'=\epsilon(\omega-\Omega m,\bz)$)
\begin{align}
&\frac{1}{2i}\int_{\inn} d\bz \left[ (\epsilon'-1) \mathbb D\,\bF_{\alpha_m}(\omega,\bz)\cdot \,{\mathbb D}^*\bF_{\beta_m}^{*}(\omega,\bz)-\mathbb D\,\bF_{\alpha_m}(\omega,\bz)\cdot (\epsilon'^*-1)\,{\mathbb D}^*\bF_{\beta_m}^{*}(\omega,\bz)\right] \nonumber \\
=&\frac{1}{2i}\int_{\inn} d\bz \left[ \left(\tilde {\mathbb D}(\epsilon' -1)\mathbb D\,\bF_{\alpha_m}(\omega,\bz)\right)\cdot \bF_{\beta_m}^{*}(\omega,\bz)-\bF_{\alpha_m}(\omega,\bz)\cdot \tilde{\mathbb D}^*(\epsilon'^*-1)\,{\mathbb D}^*\bF_{\beta_m}^{*}(\omega,\bz)\right] \nonumber \\
=&\frac{c^2}{2i\omega^2}\int_{\inn} d\bz \left[ \left(\nabla\times\nabla \times\bF_{\alpha_m}(\omega,\bz)\right)\cdot \bF_{\beta_m}^{*}(\omega,\bz)-\bF_{\alpha_m}(\omega,\bz)\cdot \nabla\times\nabla \times\bF_{\beta_m}^{*}(\omega,\bz)\right] \nonumber \\
=&\frac{c^2}{2i\omega^2}\oint d{\mathbf\Sigma}\cdot \left[ \left(\nabla \times\bF_{\alpha_m}(\omega,\bz)\right)\times \bF_{\beta_m}^{*}(\omega,\bz)+\bF_{\alpha_m}(\omega,\bz)\times \nabla \times\bF_{\beta_m}^{*}(\omega,\bz)\right],
\end{align}
where in the step from the second to the third line, we have used the EM equation, Eq.~(\ref{Eq: EM eqn moving}).
Using the continuity equations, the last line gives
\begin{align}
  \int_{\inn} d\bz \, \mathbb D\,\bF_{\alpha_m}(\omega,\bz)\cdot \im \epsilon(\omega-\Omega m, \bz)\,{\mathbb D}^*\bF_{\beta_m}^{*}(\omega,\bz)= \frac{c^2}{\omega^2}\, \delta_{\alpha_m \beta_m} \left(1-|S_{\bar\alpha_{m}}|^2\right),
\end{align}
 which is an analog of Eq.~(\ref{Eq: Integral over F: static}) for moving objects.

Therefore the corresponding EM-field correlation function becomes
\begin{align}\label{Eq: <EE> in moving}
  \langle \bE (\omega,\bx) \otimes \bE^* (\omega,\by)\rangle_{\innf}=  &\frac{\omega^2}{4c^2}\sum_{\alpha_m}a_{T}(\omega-\Omega m) \left(1-|S_{\alpha_m}|^2\right) \bE^{\out}_{\alpha_m}(\omega,\bx)\otimes \bE^{\out*}_{\alpha_m}(\omega,\by),
\end{align}
 The total radiation power can be computed via the Poynting vector to obtain
\begin{align}
  P&= \frac{1}{4\pi}\int_{0}^{\infty} d\omega \, \omega \,\sum_{\alpha_m} \left(a_{T}(\omega-\Omega m)-a_{T_0}(\omega)\right)\,  (1- |S_{\alpha_m}(\omega)|^2)\nonumber\\
  &=\int_{0}^{\infty} \frac{d\omega }{2\pi}\, \hbar \omega \,\sum_{\alpha_m} \left(n_{T}(\omega-\Omega m)-n_{T_0}(\omega)\right)\,  (1- |S_{\alpha_m}(\omega)|^2).
\end{align}
At zero temperature everywhere, $n_{0}(\omega-\Omega m)-n_{0}(\omega)=-\Theta(\Omega m -\omega)$ where $\Theta$ is the Heaviside function ($m>0$); so the radiation power in the $\alpha_m$ channel is given by
\begin{align}
  \label{em radiation}
  P_{\alpha_m}&=  \int_{0}^{\Omega m} \frac{d\omega }{2\pi}\, \hbar \omega \, (|S_{\alpha_m}(\omega)|^2-1),  \hskip .2in (m>0)
\end{align}
which in turn implies that the number of photons per unit time radiated in this channel is
\begin{align}
\frac{d{\cal N}_{\alpha_m}}{d\omega}&= \Theta(\Omega m-\omega)\, (|S_{\alpha_m}(\omega)|^2-1), \hskip .2in (m>0).
\end{align}

\section{Vacuum friction on a rotating object}
\subsection{sphere}
For a rotating object, we have to solve a complicated equation---Eq.~(\ref{rel em inside sln})---but in the lowest order we can neglect the explicit dependence on velocity while boosting the argument of the dielectric function due to rotation, {\it i.e.} $\epsilon(\omega)\to \epsilon(\omega-\Omega m)$.
For small angular velocity $\Omega$ (and at zero temperature), only frequencies within the range $0<\omega<\Omega$ contribute. In the same limit, the scattering from a sphere can be described via polarizability. We assume a non-magnetic object and so the electric polarization gives the leading contribution to the scattering amplitude, or the $T$-matrix,
\begin{equation}
  T_{1mE}(\omega)=i \frac{2\omega^3 }{3c^3}\, \alpha(\omega-\Omega m),
\end{equation}
where $\alpha(\omega)$ is the polarizability of a small sphere which depends solely on the dielectric function $\epsilon(\omega)$. Note that, according to Eq.~(\ref{em radiation}), only $m=1$ (and not $m=0,-1$) contributes to the radiation, for which the scattering matrix is
\begin{equation}
  S_{11E}=1+2T_{11E}(\omega)=1+i \frac{4}{3}\frac{\omega^3}{c^3} \alpha(\omega-\Omega).
\end{equation}
The spontaneous radiation to the vacuum is then
\begin{align}
  P&=\int_{0}^{\Omega} \frac{d\omega}{2\pi}\, \hbar \omega\left(|S_{11E}|^2-1\right) \nonumber \\
   &\approx\frac{4\hbar}{3\pi c^3}\int_{0}^{\Omega} d\omega \,\omega^4 \, \left(-\im \alpha(\omega-\Omega)\right),
\end{align}
where we have kept only the leading term in powers of frequency.
For a spherical particle of radius $R$, the polarizability is
$
\alpha(\omega)=\frac{\epsilon(\omega)-1}{\epsilon(\omega)+2} R^3,
$
hence
\begin{align}
  P=\frac{4\hbar R^3}{3\pi c^3}\int_{0}^{\Omega} d\omega \,\omega^4 \, \left|\im \frac{\epsilon(\omega-\Omega)-1}{\epsilon(\omega-\Omega)+2}\right|.
\end{align}
\subsection{cylinder}
For a cylinder, the $T$-matrices are more complicated; in fact, the two polarizations mix upon scattering from a cylindrical object. In the limit of a thin cylinder, and for $m=1$, we find ($\Omega R/c, \epsilon \, \Omega R/c \ll 1$)
\begin{align}
  T_{1k_zMM}(\omega)&= \frac{i\pi}{4} \frac{\epsilon(\omega-\Omega)-1}{\epsilon(\omega-\Omega)+1} \, \frac{\omega^2 }{c^2} R^2,\nonumber\\
  T_{1k_zEE}(\omega)&= \frac{i\pi}{4} \frac{\epsilon(\omega-\Omega)-1}{\epsilon(\omega-\Omega)+1} \, k_z^2 R^2, \nonumber\\
  T_{1k_zEM}(\omega)=T_{1k_zME}(\omega)&= \frac{i\pi}{4} \frac{\epsilon(\omega-\Omega)-1}{\epsilon(\omega-\Omega)+1} \frac{\omega k_z }{c} R^2.
\end{align}
The radiation power in the $m=1$ channel is then given by
\begin{align}
  P=\int_{0}^{\Omega} \frac{d\omega}{2\pi}\, \hbar \omega \int_{-\omega/c}^{\omega/c} \frac{L dk_z}{2\pi}\sum_{P,P'\in\{M, E\}} \left( \left|\delta^{PP'}+2\, T^{PP'}_{1k_z}(\omega)\right|^2-\delta^{PP'} \right).
\end{align}
With the approximations made, we should keep only the linear terms in $T$:
\begin{align}
  P&=\frac{2 \hbar L}{\pi}\int_{0}^{\Omega} d\omega\, \omega\int_{-\omega/c}^{\omega/c} \frac{dk_z}{2\pi} \left(\re T_{1k_zMM}(\omega)+\re T_{1k_zEE}(\omega)\right)\nonumber\\
   &=\frac {2 \hbar L R^2}{3\pi c^3}\int_{0}^{\Omega} d\omega \, \omega^4 \left|\im \frac{\epsilon(\omega-\Omega)-1}{\epsilon(\omega-\Omega)+1}\right|.
\end{align}

\section{A test object in the presence of a rotating body}
We consider the interaction of the radiation field from a rotating body with a test object. The overall EM-field correlation function is given by the sum of Eqs.~(\ref{Eq: <EE> out moving}) and (\ref{Eq: <EE> in moving})
\begin{equation}
    \langle \bE\otimes\bE^*\rangle=\langle \bE\otimes\bE^*\rangle_{\innf}+\langle \bE\otimes\bE^*\rangle_{\outf}.
\end{equation}
In the limit of zero temperature both in the object and the environment, we have
\begin{equation}
    \langle \bE \otimes \bE^* \rangle =   \langle \bE \otimes \bE^* \rangle_{\rm non-rad}+ \langle \bE \otimes \bE^* \rangle_{\rm rad} \,,
\end{equation}
where we have broken up the EM-field correlation function into radiation (due to propagating photons) and non-radiation (due to zero-point fluctuations) parts:
\begin{align}
  \langle \bE (\omega, \bx)\otimes \bE^*(\omega,\by) \rangle_{\rm non-rad}= \,\sgn(\omega) \im \mathbb G(\omega,\bx,\by),
\end{align}
where $\mathbb G$ is the Green's function in the presence of a rotating object (analog of Eq.~(\ref{Eq: G out-out})) consistent with the fluctuation-dissipation theorem, and
\begin{align} \label{Eq: G-rad}
  \langle \bE (\omega, \bx)\otimes \bE^*(\omega,\by) \rangle_{\rm rad}=\frac{\hbar\omega^2}{2c^2} \sum_{\alpha_m}\Theta(\Omega m-\omega) \, \left(|S_{\alpha_m}|^2-1\right) \bE^{\out}_{\alpha_m}(\omega,\bx)\otimes \bE^{\out*}_{\alpha_m}(\omega,\by).
\end{align}
The ``non-radiation'' correlation function resembles the equilibrium result. In fact, one can immediately see that corresponding Poynting vector is zero since the the correlation function is real. On the other hand, the ``radiation'' correlation function contributes to Poynting vector but the contribution comes from a frequency window $0<\omega<\Omega m$, in harmony with the the results of the previous sections.

To find the interaction with a \emph{test }object, we only consider the latter correlation function for two reasons. First the force due to ``non-radiation'' fluctuations typically falls off very rapidly with the separation distance, while the radiation pressure exerts a force falling off slowly. Second, since we consider the test object to be spherical, only the radiation term can give rise to a tangential force or torque. This is because the ``non-radiation'' fluctuations give rise to a potential energy depending only on the separation distance.

The radiation from a (non-magnetic) rotating sphere is dominated by the lowest (electric) partial wave $(l=1,m=1,P=E)$, so we find
\begin{align}
  \langle \bE (\omega, \bx)\otimes \bE^*(\omega,\by) \rangle_{\rm rad}= \frac{\hbar\omega^2}{2c^2}  \,\Theta(\Omega-\omega)\left(|S_{11E}|^2-1\right) \bE^{\out}_{11E}(\omega,\bx)\otimes \bE^{\out*}_{11E}(\omega,\by).
\end{align}
The partial waves in Eq.~(\ref{Eq: G-rad}) are defined in spherical basis as
\begin{align}\label{Eq: spherical fns}
  &\bE^{\out}_{lmM}(\omega,\bx)=\frac{\sqrt{\omega/c}}{\sqrt{l(l+1)}} \nabla \times \, h_l^{(1)}\left(\frac{\omega r}{c}\right)Y_{lm}(\theta, \phi) \bx, \nonumber \\
  &\bE^{\out}_{lmE}(\omega,\bx)=-i\frac{\sqrt{c/\omega}}{\sqrt{l(l+1)}} \nabla \times \nabla \times \, h_l^{(1)}\left(\frac{\omega r}{c}\right)Y_{lm}(\theta, \phi) \bx,
\end{align}
where $Y_{lm}$ is the usual spherical harmonic function. Note that the normalization constants are chosen to conform with Eq.~(\ref{Eq: identity}).

In order to find the scattering from the second object, we expand the EM field around its origin located at a separation $d$ {on} the $x$ axis. To the lowest order in frequency, we have
\begin{align}
  \bE^{\out}_{11E}(\omega,\bx)=\mathcal U_{11E, 11E}   \bE^{\reg}_{11E}(\omega,\tilde \bx) +\mathcal U_{10M, 11E}   \bE^{\reg}_{10M}(\omega,\tilde \bx)+\cdots,
\end{align}
where $\tilde \bx$ is defined with respect to the new origin. The {\it regular} functions, denoted by ``reg'', are defined by replacing the spherical Hankel function $h^{(1)}_l$ in Eq.~(\ref{Eq: spherical fns}) by the spherical Bessel function $j_l$. The {\it translation }matrices are given by
\begin{align}
   \mathcal U_{11E, 11E}= h^{(1)}_0\left(\frac{\omega d}{c}\right), \hskip .3in \mathcal U_{10M, 11E}= \frac{\sqrt{2}\omega d}{4c}h^{(1)}_0\left(\frac{\omega d}{c}\right).
\end{align}
Next we consider the scattering from the test object:
\begin{equation}
  \bE^{\reg}_{lmP}(\omega,\tilde \bx)\rightarrow \frac{1}{2}\left(\bE^{\inn}_{lmP}(\omega,\tilde\bx) +\mathfrak{S}_{lmP}\bE^{\out}_{lmP}(\omega,\tilde\bx)\right),
\end{equation}
where $\mathfrak{S}_{lmP}$ is the scattering matrix of the test object. Putting these equations together, one can find the EM-field correlation function upon one scattering from the test object
\begin{align}
  \langle \bE (\omega, \tilde \bx)\otimes &\bE^*(\omega,\tilde \by) \rangle_{\rm rad}=\frac{\hbar \omega^2}{2c^2}  \,\Theta(\Omega-\omega)\left(|S_{11E}|^2-1\right) \times\nonumber \\ &\times\frac{1}{4}\left[\mathcal U_{11E, 11E}\left(\bE^{\inn}_{11E}(\omega,\tilde\bx) +\mathfrak{S}_{11E}\bE^{\out}_{11E}(\omega,\tilde\bx)\right)+\mathcal U_{10M, 11E}\left(\bE^{\inn}_{10M}(\omega,\tilde\bx) +\mathfrak{S}_{10M}\bE^{\out}_{10M}(\omega,\tilde\bx)\right)\right]\nonumber \\
  &\otimes \overline{\left[\mathcal U_{11E, 11E}\left(\bE^{\inn}_{11E}(\omega,\tilde\by) +\mathfrak{S}_{11E}\bE^{\out}_{11E}(\omega,\tilde\by)\right)+\mathcal U_{10M, 11E}\left(\bE^{\inn}_{10M}(\omega,\tilde\by) +\mathfrak{S}_{10M}\bE^{\out}_{10M}(\omega,\tilde\by)\right)\right]},
\end{align}
where we have suppressed the contributions from higher partial waves since they come in higher orders in frequency. (The overbar notation denotes complex conjugation.)

We compute the torque and the shear force. For the torque, the partial waves $11E$ and $10M$ decouple, but the latter does not contribute since its azimuthal index, $m$, is zero. The torque is then
\begin{align}
  M&= \frac{\hbar }{4} \int_{0}^{\Omega} \frac{d\omega}{2\pi} \, (|S_{11E}|^2-1) \left|{\cal U}_{11E,11E}\right|^2(1-|\mathfrak{S}_{11E}|^2), \nonumber \\
   &=\frac{\hbar c^2}{8\pi d^2} \int_{0}^{\Omega} d\omega \, \frac{1}{\omega^2} (|S_{11E}|^2-1)(1-|\mathfrak{S}_{11E}|^2).
\end{align}
The force is more complicated since the two partial waves mix, and one has to find their overlap with the Maxwell stress tensor. A lengthy, though straightforward, calculation leads to
\begin{align}
      F_y&=    \frac{\hbar}{2}\int_{0}^{\Omega} \frac{d\omega}{2\pi} \frac{\omega^2}{c^2} \,(|S_{11E}|^2-1) \,\, \mathcal U_{10M, 11E} \mathcal U_{11E, 11E} \,\mathcal T_{10M,11E} \,\re(-1+\overline{\mathfrak{S}_{10M}}\mathfrak{S}_{11E}),
\end{align}
where $\mathcal T_{10M,11E}$ characterizes the stress tensor sandwiched between the two partial waves, whose dependence on frequency is given by
\begin{equation}
  \mathcal T_{10M,11E} =-\frac{\pi c}{2\sqrt{2}\omega}.
\end{equation}
For a non-magnetic object, we can safely assume $\mathfrak{S}_{10M}\approx1$; its frequency dependence can be neglected compared to $\mathfrak{S}_{11E}$. The force is then
\begin{align}
F_y=\frac{\hbar}{32\pi d}\int_{0}^{\Omega} d\omega \,(|S_{11E}|^2-1) (1-\re\mathfrak{S}_{11E}).
\end{align}

\section*{Appendix A: Green's theorem}\label{appen}

The vector Green's theorem reads
\begin{equation}
  \bE_i(\bx)=\oint d\mathbf\Sigma\cdot \left[ (\nabla \times \bG_i(\bx,\bz))\times \bE (\bz) +\bG_i(\bx,\bz)\times (\nabla \times\bE(\bz)) \right],
\end{equation}
where $\bE_i$ refers to the $i$ component of the electric field, $\bG_i$ is a vector defined from the dyadic Green's function as $(\bG_i)_j=\mathbb G_{i j}$, and $\bE$ is a solution to the vector Helmholtz equation. Also note that the point $\bx$ is enclosed by the boundary of the integration. Next we choose $\bE=\bE^{\reg}_\beta$, a partial wave indexed by $\beta$, and also employ the definition of the Green's function in Eq.~(\ref{Eq: G out-out0}) to find
\begin{equation}
  (\bE^{\reg}_\beta(\bx))_i=i\sum_{\alpha} (\bE^{\reg}_\alpha(\bx))_i \oint d\mathbf\Sigma\cdot \left[ (\nabla \times \bE^{\out}_{\bar \alpha}(\bz))\times \bE^{\reg}_\beta(\bz) +\bE^{\out}_{\bar\alpha}(\bz)\times (\nabla \times\bE^{\reg}_{\beta}(\bz)) \right].
\end{equation}
The (vector) wave-functions $\bE^{\reg}_{\alpha}$ constitute a complete set, hence
\begin{equation}
  i \oint d\mathbf\Sigma\cdot \left[ (\nabla \times \bE^{\out}_{\bar \alpha}(\bz))\times \bE^{\reg}_\beta(\bz) +\bE^{\out}_{\bar\alpha}(\bz)\times (\nabla \times\bE^{\reg}_{\beta}(\bz)) \right]=\delta_{\alpha \beta}.
\end{equation}
Using the definition of the ``regular'' wave-functions, we arrive at Eq.~(\ref{Eq: identity}).


\end{document}